\documentclass[showpacs,preprintnumbers,amsmath,amssymb]{revtex4}
\usepackage{amsmath,amssymb,graphics,epsfig,subfigure}
\usepackage{color}

\usepackage[normalem]{ulem}

\begin{document}
\renewcommand{\baselinestretch}{1.3}
\title{Constraining rotating black hole via curvature radius with observations of M87*}

\author{Shao-Wen Wei$^{a,b,c}$\footnote{weishw@lzu.edu.cn, corresponding author},
Yuan-Chuan Zou$^{d}$\footnote{zouyc@hust.edu.cn}}

\affiliation{$^{a}$Lanzhou Center for Theoretical Physics, Key Laboratory of Theoretical Physics of Gansu Province, School of Physical Science and Technology, Lanzhou University, Lanzhou 730000, People's Republic of China,\\
 $^{b}$Institute of Theoretical Physics $\&$ Research Center of Gravitation,
Lanzhou University, Lanzhou 730000, People's Republic of China,\\
$^{c}$Academy of Plateau Science and Sustainability, Qinghai Normal University, Xining 810016, People's Republic of China\\
$^{d}$Department of Astronomy, School of Physics, Huazhong University of Science and Technology, Wuhan 430074, People's Republic of China}

\begin{abstract}
Combined with the observation of M87*, shadow has gradually became a promising test of the black hole nature. Recently, EHT collaboration gave new constraints on the shadow size at 68\% confidence levels. In this work, we consider the new constrains on the black hole spin and charge for the Kerr and Kerr-Newmann black holes via the local curvature radius. For the Kerr black holes, the cases with high spin and large inclination angle are ruled out. For the Kerr-Newmann black holes, two new characteristic constrained patterns for low and high black hole spins are given. Near extremal black holes are always excluded unless for the high spin and low inclination angle. Moreover, we find that low black hole spin and charge can pass these constraints as expected. These results suggest that this approach of curvature radius is effective on constraining black hole parameters. We expect the local concept of the curvature radius will play a more important role on the study of the black hole shadow in the further.
\end{abstract}



\maketitle

\section{Introduction}
\label{secIntroduction}

A couple of years ago, Event Horizon Telescope (EHT) Collaboration released the first image of the supermassive black hole M87* \cite{Akiyama1}. The results provided us with a promising test to the strong gravity regime near a black hole horizon. By modeling M87* with a Kerr black hole solution, EHT found that the observations of the black hole shadow are well consistent with the prediction of general relativity. However, modified gravities are not completely ruled out by these observations. How to test the nature of the black hole is also a challenge. Aiming on these purposes, many works have been done, see Refs. \cite{Moffat,Shaikha,Mingzhi,Freeseaa,Kimet,Tao,Cheng,Jie,Hui,ShaoWen} for examples. In particular, with future observations, like the Next Generation Very Large Array \cite{Hughes}, the Thirty Meter Telescope \cite{Sanders}, and the BlackHoleCam \cite{Goddi}, we will have good opportunity to peek into the strong gravity regime.

The study of the black hole shadow can be traced back to the work \cite{Synge}, where Synge studied the observed angular radius of a Schwarzschild black hole. Later studies can also be found in Refs. \cite{Luminet,Bardeen,Chandrasekhar}. The shadows cast by spherically symmetric and axial symmetric black holes are well understood. Now it is believed that the formation of the shadow is mainly caused by the existence of the photon sphere or light ring rather than the horizons. For a spherically symmetric black hole, its shadow is a perfect circle. However when a black hole gets rotation, its shadow will be elongated in the direction of the rotating axis due to spacetime dragging effect. Also, it is found that the size and deformation of the shadow vary with different black holes. This property allows one to determine the black hole parameters from the shadow.

After obtaining a shadow shape, how to determine the black hole parameters from the observation is an urgent problem. The key of it is constructing the observables which will bridge the theory and the observation. In Ref. \cite{Hioki}, Hioki and Maeda first considered this issue. Several observables were proposed and extensively used on testing black hole parameters. The shadow size and deformation can be well described by the radius of the reference circle and the distortion parameter. The area and perimeter of the shadow are also examined. Since these observables are related with the global information of the shadow, we call them global observables. They are very useful on determining the black hole parameter once the complete shadow shape is obtained \cite{Amarilla2,Johannsen,Ghasemi,Bambi,Amarilla,Nedkova,Wei,Bambi3,Atamurotov,WWei,AmirAhmedov,Songbai,Balendra,
Tsukamoto2,hou,Cunha,Cunha2,Tsupko,Perlick,Younsi,Akashaa,Abdujabbarov,
Konoplyab,Vagnozzib,Banerjeeb,Lua,Fengb,Renb,Jusufi}.

Very recently, EHT Collaboration imposed a constraint on the black hole charge via the characteristic areal-radius relating with the area of the shadow \cite{Kocherlakota}. By making use of the 2017 EHT observations of M87*, they suggested that the characteristic areal-radius must fall in a range (4.31$M$, 6.08$M$) at 68\% confidence levels. The results indicate that highly charged dilaton black hole has been ruled out by the observation of M87*. Moreover, a considerable region of the space of parameters for the doubly-charged dilaton and the Sen black holes is also excluded. This study shows that the global observables are quite powerful on the study of black hole shadow. This observation was also used to constrain the tidal charge of the supermassive black hole M87* \cite{Zakharovb}.

Differently, local observables will give us a novel point of view on the shadow. It will be useful when we only have a part image of the shadow in high resolution rather than a complete one. Considering this advantage, we first proposed a local observable, the curvature radius, for the shadow in Ref. \cite{WeiLiu}. The curvature radius varies along the shadow curve. For the Kerr black hole, the curvature was calculated, which can be expressed in a compact form in terms of the black hole spin, inclination angle, and shadow curve parameter. After excluding the $Z_2$ symmetry of the shadow, we found that there exist one minimum and one maximum for each shadow. Employing these two values, we developed a new approach to determine the black hole parameters. Subsequently, we examine the curvature radius for the Kerr black hole in a more general approach \cite{WeiLiu2}. Comparing with these global observables, it implies that less information is available for the local observables to exactly determine the black hole.

On the other hand, the minimum and maximum of the curvature radius provide us with natural lower and upper bounds for the shadow size. Basing on this idea, we in this work will constrain the Kerr and Kerr-Newmann (KN) black holes with the observations of M87* that the radius of shadow size is in the range (4.31$M$, 6.08$M$). The paper is organized as follows. In Sec. \ref{null}, we firstly analyze the local curvature radius for the Kerr black hole. Then we constrain the black hole spin by the minimum and maximum of the curvature radius. In Sec. \ref{KN}, we calculate the curvature radius for the KN black hole. Then the black hole spin and charge are constrained. At last, we summarize our result in Sec. \ref{Conclusion}.

\section{Constraints for Kerr black hole}
\label{null}

Here, we would like to give a brief review to the Kerr black hole and its shadow, and then constrain the black hole spin.

\subsection{Kerr black hole and shadow}
In Boyer-Lindquist coordinates, the rotating Kerr black hole solution is given by
\begin{eqnarray}
 ds^{2}=-\left(1-\frac{2Mr}{\rho^{2}}\right)dt^{2}+\frac{\rho^{2}}{\Delta}dr^{2}+\rho^{2}d\theta^{2}
 -\frac{4Mra\sin^{2}\theta}{\rho^{2}}dtd\phi+\frac{[(r^{2}+a^{2})^{2}-\Delta a^{2}\sin^{2}\theta]\sin^{2}\theta}{\rho^{2}}d\phi^{2},\label{metric}
\end{eqnarray}
where the metric functions read
\begin{eqnarray}
 \Delta=r^{2}-2Mr+a^{2},\quad
 \rho^{2}=r^{2}+a^{2}\cos^{2}\theta.
\end{eqnarray}
The parameters $M$ and $a$ are, respectively, the mass and spin of the black hole. The horizons located at $\Delta(r_\pm)=0$ are
\begin{eqnarray}
 r_{\pm}=M\pm \sqrt{M^{2}-a^{2}}.
\end{eqnarray}
Obviously, when $|a|<M$, there are one event horizon $r_{+}$ and one Cauchy horizon $r_{-}$. These two horizons coincide at $|a|=M$ for the extremal black hole. A naked singularity will be present when $|a|>M$.

\begin{figure}
\center{\subfigure[]{\label{ShadowKa}
\includegraphics[width=7cm]{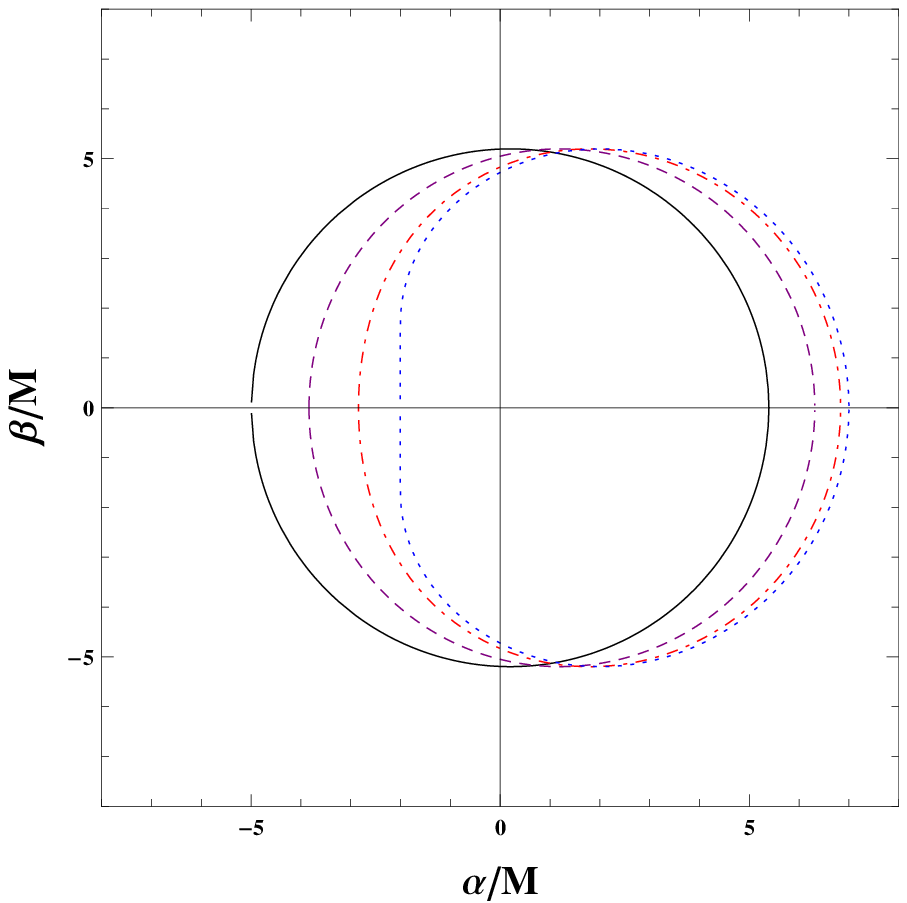}}
\subfigure[]{\label{ShadowKTh}
\includegraphics[width=7cm]{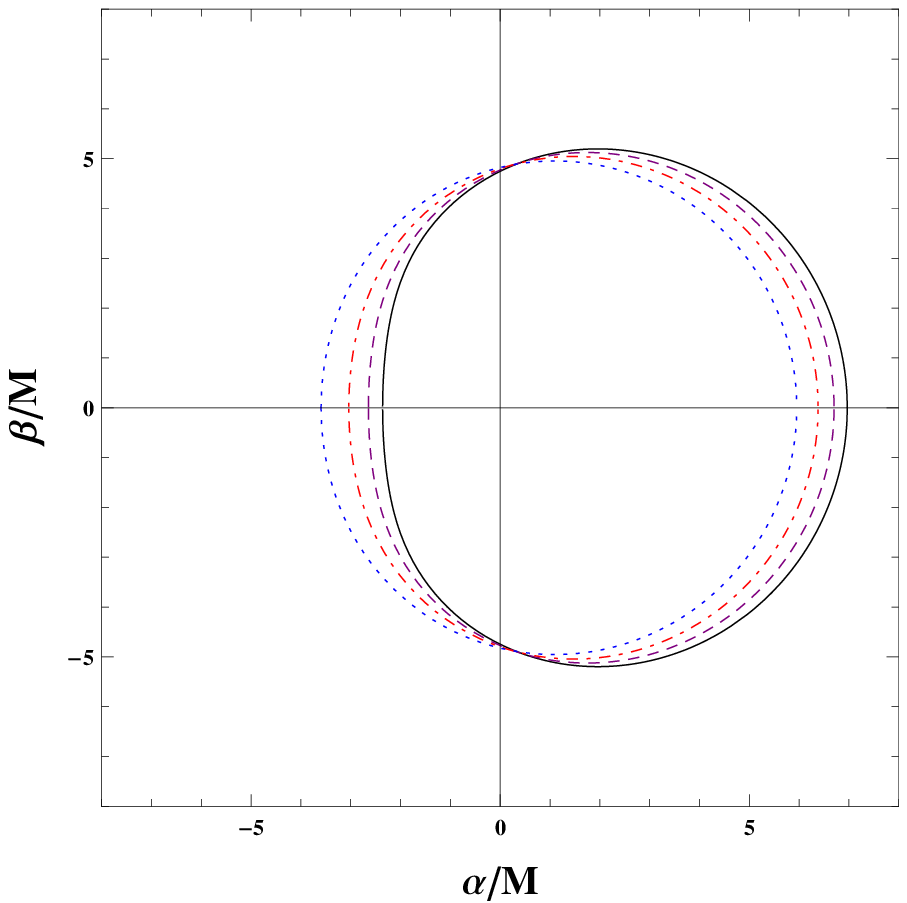}}}
\caption{Shadows cast by Kerr black holes. (a) $\theta_{0}=\frac{\pi}{2}$ with $a/M$=0 (black solid line), 0.6 (dashed purple line), 0.9 (red dot dashed line), and 1 (blue dot line) from left to right. (b) $a/M$=0.95 with $\theta_0$=$\frac{\pi}{2}$ (black solid line), $\frac{\pi}{3}$ (dashed purple line), $\frac{\pi}{4}$ (red dot dashed line), $\frac{\pi}{6}$ (blue dot line) from right to left.}\label{pShadowKTh}
\end{figure}

For each geodesics, there are three conserved quantities $l$, $E$, and $\mathcal{Q}$, which are related with two Killing vector fields and one Killing-Yano tensor field in this spacetime. As well known, $l$ and $E$ are the angular momentum and energy for the test particle, and $\mathcal{Q}$ is the Carter constant. It is convenient to introduce two new conserved parameters $\xi$ and $\eta$
\begin{eqnarray}
 \xi=\frac{l}{E}, \quad \eta=\frac{\mathcal{Q}}{E^{2}},
\end{eqnarray}
and this is equivalent to set $E=1$. So the parameters $\xi$ and $\eta$ correspond to the angular momentum and Carter constant, respectively. With the help of these three conserved quantities, the null geodesics in Kerr spacetime can be obtained
\begin{eqnarray}
 \rho^{2}\dot{t}&=&\frac{(r^{2}+a^{2})^{2}-2Mra\xi}{\Delta}-a^{2}\sin^{2}\theta,\\
 \rho^{2}\dot{\mathcal{R}}&=&\pm\sqrt{\mathcal{R}},\\
 \rho^{2}\dot{\theta}&=&\pm\sqrt{\Theta},\\
 \rho^{2}\dot{\phi}&=&\frac{2Mar+\xi\csc^{2}\theta(\rho^{2}-2Mr)}{\Delta},
\end{eqnarray}
where the dot denotes the derivative with respect to the affine parameter. The functions $\mathcal{R}$ and $\Theta$ are given by
\begin{eqnarray}
 \mathcal{R}&=&(r^{2}+a^{2}-a\xi)^{2}-\Delta(\eta+(a-\xi)^{2}),\\
 \Theta&=&\eta+(a-\xi)^{2}-(a\sin\theta-\xi\csc\theta)^{2}.
\end{eqnarray}
By making use of the null geodesics, the black hole shadow can be formulated. Here we suppose that the light sources are located at infinity and distributed uniformly in all directions, and the observer also located at infinity while with the inclination angle $\theta_{0}$. In order to clearly describe the shapes for the shadow, it is convenient to introduce the celestial coordinates. For a locally nonrotating observer, the celestial coordinates \cite{Bardeen,Chandrasekhar} are
\begin{eqnarray}
 \alpha &=& -\lim_{r\rightarrow\infty}\frac{rp_{\phi}}{p_{t}}\bigg|_{\theta\rightarrow\theta_{0}},\\
 \beta  &=&  \pm\lim_{r\rightarrow\infty}\frac{rp_{\theta}}{p_{t}}\bigg|_{\theta\rightarrow\theta_{0}},
\end{eqnarray}
where $(r, \theta_{0})$ denotes the coordinate of the observer. From the null geodesics, $\alpha$ and $\beta$ can be explicitly expressed in terms of $\xi$, $\eta$, and $\theta_{0}$
\begin{eqnarray}
 \alpha&=&-\xi\csc\theta_{0},\label{oe}\\
 \beta&=&\pm\sqrt{\eta+a^{2}\cos^{2}\theta_{0}-\xi^{2}\cot^{2}\theta_{0}}.\label{tho}
\end{eqnarray}
with
\begin{eqnarray}
 \xi&=&\frac{r_{0}^2+a^2}{a}-\frac{2r_{0}\Delta(r_{0})}{a(r_{0}-M)},\label{xx}\\
 \eta&=&\frac{r_{0}^3\left(4a^2M-r_{0}(r_{0}-3M)^2\right)}{a^{2}(r_{0}-M)^{2}}.\label{ee}
\end{eqnarray}
Here $r_{0}$ is in a certain range, which depends on the black hole spin $a$ and the inclination angle $\theta_{0}$. If the observer is located in the equatorial plane, $\theta_{0}=\pi/2$, the celestial coordinates will be simplified as $\alpha=-\xi$ and $\beta=\pm\sqrt{\eta}$. Meanwhile, we have $r_{0}\in(r_{1},\;r_{2})$ with
\begin{eqnarray}
 r_{1,2}=2M\left(1+\cos\left(\frac{2}{3}\arccos\left(\mp\frac{a}{M}\right)\right)\right).
 \label{rr}
\end{eqnarray}
Here we list the shadows cast by Kerr black holes in Fig. \ref{pShadowKTh}. From Fig. \ref{ShadowKa}, one can find that for $\theta_{0}$=$\frac{\pi}{2}$, the shape is almost a standard circle for lower black hole spin, while is obviously deformed for higher spin. For the fixed spin $a$=0.95, see Fig. \ref{ShadowKTh}, the shapes is shifted toward the left with the decrease of $\theta_0$.

%%%%%%%%%%%%%%%%%%%%%%%%%%%
%%%%%%%%%%%%%%%%%%%%%%%%%%%

\subsection{Curvature radius and constraints}

Curvature radius is a local concept from the differential geometry, which is powerful on describing one-dimensional curve. Considering that the boundary of the shadow is a one-dimensional closed loop and each shadow has one characteristic loop, we firstly introduce the curvature radius into the study of the black hole shadow \cite{WeiLiu}. Employing such novel concept, we have studied how to test the black hole nature from the shadow, and which seems to be a valuable approach.

Recently, In Ref. \cite{Kocherlakota}, EHT Collaboration adopted a characteristic areal-radius $r_{\rm sh,A}$ of the shadow curve to approximately measures the shadow size, and suggest a lower and upper bounds, i.e.,
\begin{eqnarray}
 4.31M\leq r_{\rm sh,A}\leq6.08M \label{diddd}
\end{eqnarray}
at 68\% confidence levels via the observation of M87* in 2017. The circle with the radius $r_{\rm sh,A}$ has the same area with the black hole shadow. So, the radius $r_{\rm sh,A}$ gives us a global information of the shadow. Instead, the curvature radius $R$ is a local concept. For a fixing shadow, it takes different values at different boundary points. From the local point of view, we can transfer the constraint (\ref{diddd}) to
\begin{eqnarray}
 4.31M\leq R \leq6.08M \label{cons}
\end{eqnarray}
This might be a good constraint when one considering the shadow cast by rotating black holes. In the following section, we will constrain the black hole parameter by using (\ref{cons}).

For the shadow cast by Kerr black hole, the curvature radius is given by \cite{WeiLiu}
\begin{eqnarray}
 R=\frac{64M^{1/2}(r_{0}^{3}-a^{2}r_{0}\cos^{2}\theta_{0})^{3/2}\left(r_{0}(r_{0}^{2}-3Mr_{0}+3M^{2})-a^{2}M^{2}\right)}
   {(r_{0}-M)^{3}\left(3(8r_{0}^{4}-a^{4}-8a^{2}r_{0}^{2})-4a^{2}(6r_{0}^{2}+a^{2})\cos(2\theta_{0})
    -a^{4}\cos(4\theta_{0})\right)}.
\label{Rad00}
\end{eqnarray}
In Ref. \cite{WeiLiu}, we studied the behavior of $R$ as a function of the line length parameter. The results indicate that there exist one maximum and one minimum of the radius for each shadow shape when we exclude the $Z_2$ symmetry.

%%%%%%%%%%%%%%%%%%%%%%%%%%%
\begin{figure}
\center{\subfigure[]{\label{RKerrTha}
\includegraphics[width=7cm]{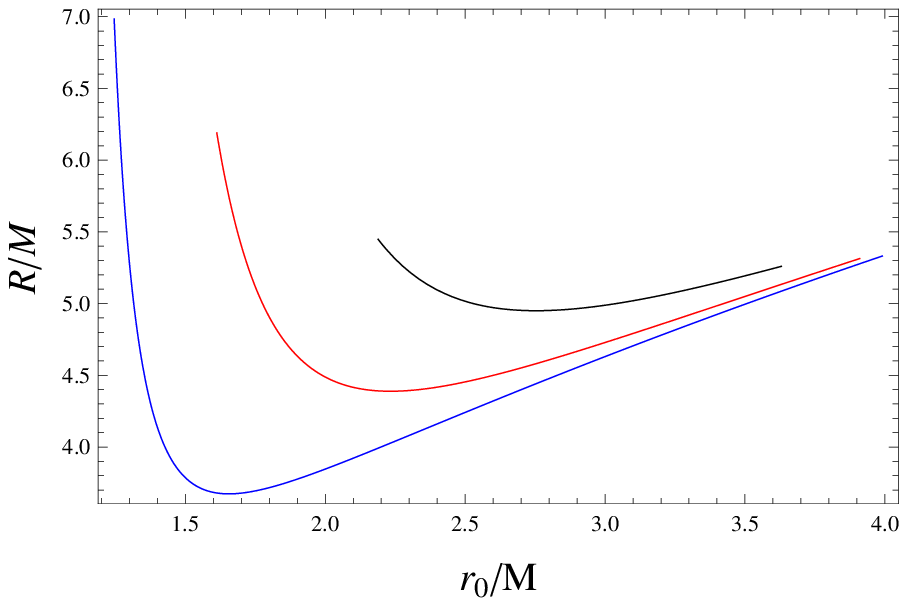}}
\subfigure[]{\label{RKerra}
\includegraphics[width=7cm]{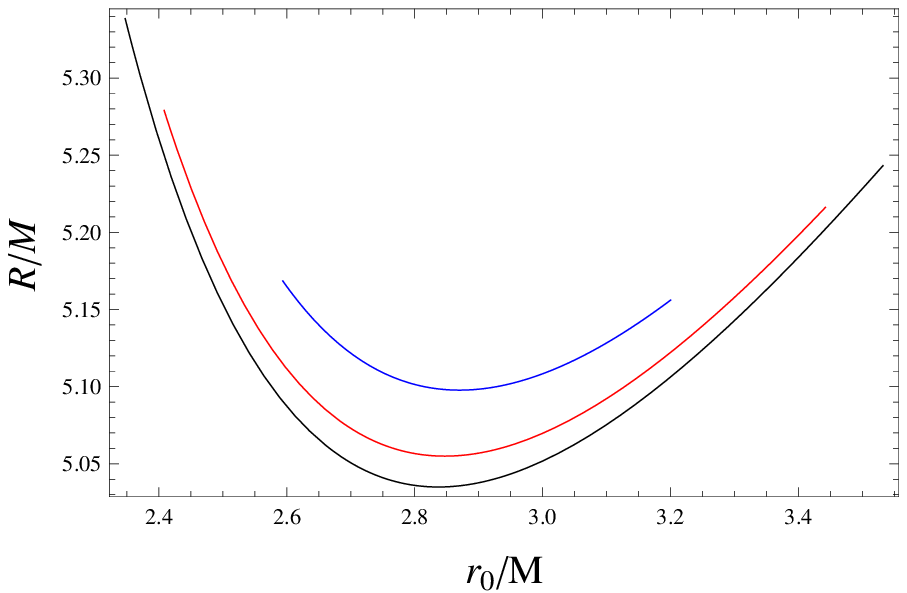}}}
\caption{Curvature radius $R$ as a function of $r_0$. (a) $\theta_0=\frac{\pi}{2}$ with black hole spin $a/M$=0.6, 0.9, and 0.99 from top to bottom. (b) $a/M$=0.5 with $\theta_0$=$\frac{\pi}{2}$, $\frac{\pi}{3}$, and $\frac{\pi}{6}$ from bottom to top.}\label{pRKerra}
\end{figure}
%%%%%%%%%%%%%%%%%%%%%%%%%%%

Here we exhibit the curvature radius with allowed $r_{0}\in(r_1, r_2)$ in Fig. \ref{pRKerra}. After a simple calculation, we find that the range of the allowed $r_0$ widens with $a$ and $\theta_0$. For $\theta_0=\pi/2$, $r_1$ and $r_2$ are given in (\ref{rr}). In particular, from these curves, we find that with the increase of $r_0$, $R$ decreases firstly and then increases. So such decreasing-increasing behavior is a characteristic feature of $R$. For each case, the local maximum $R$ are at $r_0$=$r_1$ and $r_2$. Since $R(r_1)>R(r_2)$, the maximum value occurs at the left, i.e., $R_{\rm max}=R(r_1)$. The minimum $R_{\rm min}$ locates at the well of the curve. From Fig. \ref{RKerrTha} with $\theta_0$=$\frac{\pi}{2}$, it is easy to find that $R_{\rm max}$ increases, while $R_{\rm min}$ decreases with the black hole spin $a$. For fixed $a/M$=0.5, $R$ is plotted in Fig. \ref{RKerra}, we observe that $R_{\rm max}$ increases and $R_{\rm min}$ decreases with $\theta_0$. So we can conclude that $R_{\rm max}$ increases and $R_{\rm min}$ decreases with $a$ and $\theta_0$. Eventually, $R_{\rm max}$ and $R_{\rm min}$ provide us natural lower and upper bounds of the shadow size, which can be used to constrain the black hole parameters via the observation of M87*. Then the constraint (\ref{cons}) can be further expressed as
\begin{eqnarray}
 4.31M\leq R_{\rm min}, R_{\rm max} \leq6.08M. \label{cons2}
\end{eqnarray}
Therefore, we can implement these constrains by calculating $R_{\rm max}$ and $R_{\rm min}$.

%%%%%%%%%%%%%%%%%%%%%%%%%%%
\begin{figure}
\includegraphics[width=7cm]{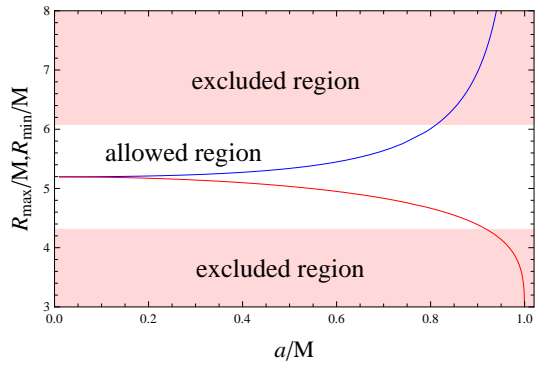}
\caption{$R_{\rm max}$ and $R_{\rm min}$ as a function of black hole spin $a$. Shadow regions are the excluded regions, where the black hole shadow size is too small or too large.}\label{ConKerr}
\end{figure}
%%%%%%%%%%%%%%%%%%%%%%%%%%%

%%%%%%%%%%%%%%%%%%%%%%%%%%%
\begin{figure}
\includegraphics[width=7cm]{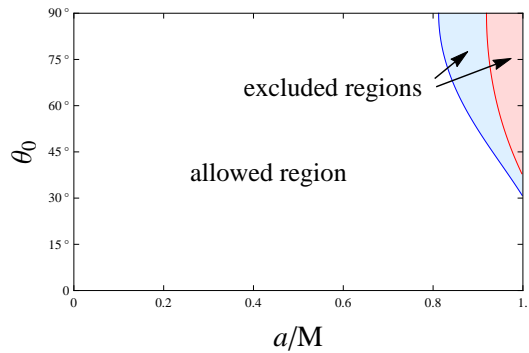}
\caption{Excluded and allowed regions in $\theta_0$-$a/M$ plane. Blue (left) and red (right) curves are the constrained spins for $R_{\rm max}$ and $R_{\rm min}$, respectively.}\label{ConKerraTh}
\end{figure}
%%%%%%%%%%%%%%%%%%%%%%%%%%%

For fixing $\theta_0$=$\frac{\pi}{2}$, we calculate $R_{\rm max}$ and $R_{\rm min}$ varying with the black hole spin $a$ in Fig. \ref{ConKerr}. The shadow regions denote the excluded regions where $R_{\rm min}<4.31M$ or $R_{\rm max}>6.08M$, otherwise, it is allowed region. From Fig. \ref{ConKerr}, we can see that when $a/M<0.8123$, $R_{\rm max}$ is smaller than 6.08$M$, and $R_{\rm min}$ is larger than 4.31$M$ for $a/M<0.9189$. If we require the constrains (\ref{cons2}) is satisfied both for $R_{\rm max}$ and $R_{\rm min}$, one must have the constrained spin
\begin{eqnarray}
 a/M\leq0.8123.
\end{eqnarray}

On the other hand, we know from Fig. \ref{ShadowKTh} that the shadow size not only depends on the black hole spin $a$, but also on the inclination angle $\theta_{0}$. So it is natural to ask if the observer deviates from the equatorial plane, how does the constrained spin behave. In order to response it, we calculate $R_{\rm max}$ and $R_{\rm min}$ and find the constrained spins for different $\theta_{0}$. The result is exhibited in Fig. \ref{ConKerraTh}. From the figure, one can find that the constrained spins for $R_{\rm max}$ and $R_{\rm min}$ decreases with $\theta_{0}$. When the observer is in the equatorial plane, it gives us a strong constraint. While when $\theta_{0}<38^{\circ}$, $R_{\rm min}$ will has no constraint on the black hole spin, and when $\theta_{0}<31^{\circ}$, $R_{\rm min}$ will also lose it constraint. Therefore, for low inclination angle $\theta_{0}$, i.e.,
\begin{eqnarray}
 \theta_{0}<31^{\circ}
\end{eqnarray}
we have no any constraint on the black hole spin, and thus the Kerr black hole can pass the observation of M87*.

\section{Constraints for Kerr-Newmann black hole}
\label{KN}

Comparing with the Kerr black hole, the charge is presented for the KN black hole. As shown in Ref. \cite{Kocherlakota}, EHT gave the constraints for the black hole charges by employing the characteristic areal-radius. Here, we would like to consider the constraints for the KN black hole combining with curvature radius and the observation of the shadow of M87*.

\subsection{Kerr-Newmann black hole and shadow}

The KN black hole share the same line element (\ref{metric}) as the Kerr black hole, while with a different metric function
\begin{eqnarray}
 \Delta=r^{2}-2Mr+a^{2}-Q^2,
\end{eqnarray}
where $Q$ is the black hole charge. Solving it, it is easy to obtain the horizons
\begin{eqnarray}
 r_{\pm}=M\pm \sqrt{M^{2}-a^{2}+Q^2}.
\end{eqnarray}
The extremal black hole with a degenerate horizon satisfies
\begin{eqnarray}
 \left(\frac{a}{M}\right)^2+\left(\frac{Q}{M}\right)^2=1.
\end{eqnarray}
When $(a/M)^2+(Q/M)^2>1$, the solution just describes a naked singularity. Here we only consider the black hole cases, which imposes a maximal bound on the black hole spin
\begin{eqnarray}
 a_{\rm max}/M=\sqrt{1-(Q/M)^2},
\end{eqnarray}
or equivalently
\begin{eqnarray}
 Q_{\rm max}/M=\sqrt{1-(a/M)^2},
\end{eqnarray}
for fixing $Q$ or $a$, respectively.

For the KN black hole, the parameters $\xi$ and $\eta$ are given by
\begin{eqnarray}
 \xi&=&\frac{r_{0}^2+a^2}{a}-\frac{2r_{0}\Delta(r_{0})}{a(r_{0}-M)},\label{xx}\\
 \eta&=&\frac{r_0^2 \left(4 a^2 \left(M
   r_0-Q^2\right)-\left( r_{0}^2-3 Mr_0+2
   Q^2\right)^2\right)}{a^2
   \left(r_0-M\right)^2}.\label{eed}
\end{eqnarray}
When the black hole charge $Q$ vanishes, it will reduce back to (\ref{ee}). However, these two celestial coordinates $\alpha$ and $\beta$ are still in the forms (\ref{oe}) and (\ref{tho}). Varying $Q$, $a$, and $\theta_0$, we will obtain the shadow shape for the KN black holes.

\subsection{Curvature radius and constraints}

After obtaining the explicit forms of the celestial coordinates $\alpha$ and $\beta$ for the KN black hole, we can calculate the curvature radius. Further, we can constrain the black hole spin and charge via the observation of M87*.

\begin{figure}
\center{\subfigure[]{\label{KNRVa}
\includegraphics[width=7cm]{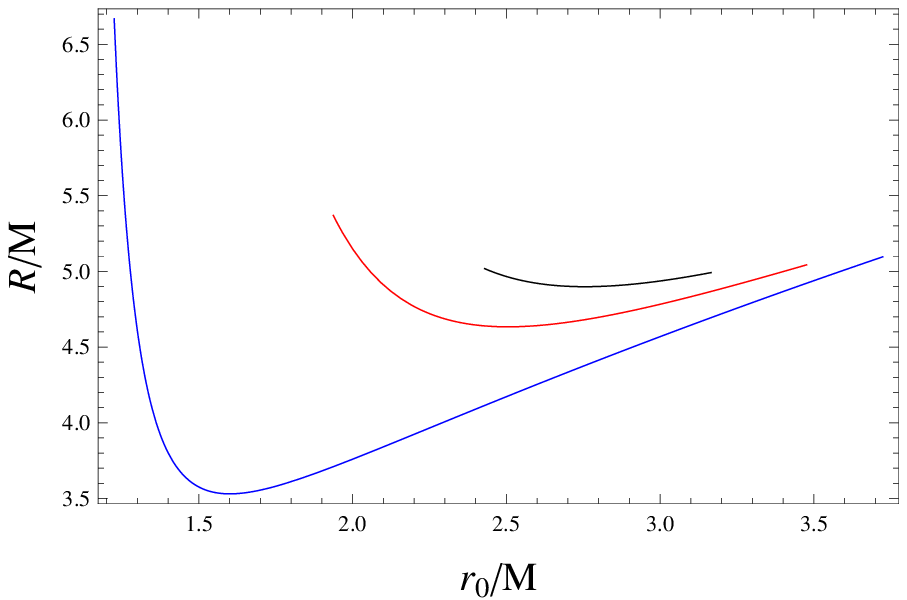}}
\subfigure[]{\label{KNRVQ}
\includegraphics[width=7cm]{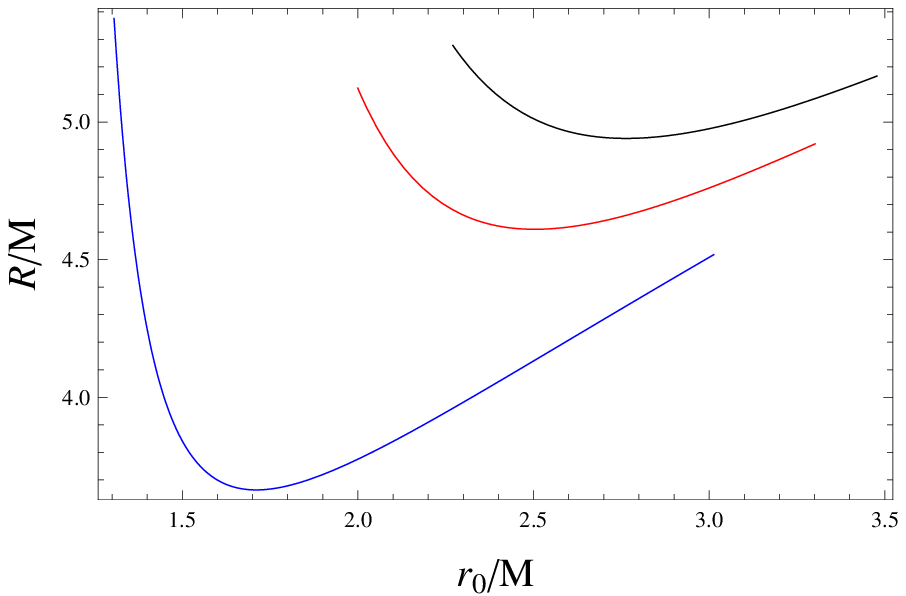}}}
\caption{Behaviors of the curvature radius $R$ for the KN black hole with $\theta_0$=$\frac{\pi}{2}$. (a) $Q/M$=0.5 with $a/M$=0.3, 0.5, and $0.99*a_{\rm max}/M$ from top to bottom. (b) $a/M$=0.5 with $Q/M$=0.3, 0.6, $0.99*Q_{\rm max}/M$ from top to bottom.}\label{pKNRVQ}
\end{figure}

We know that three non collinear points can uniquely determine one circle. In order to find the curvature radius at a certain point parameterized by $r_0$ of the shadow boundary, we need other two neighbouring points, which we parameterize them with $r_0-\epsilon$ and $r_0+\epsilon$. Then we have three points in the $\alpha$-$\beta$ plane, i.e., ($\alpha(r_0-\epsilon)$, $\beta(r_0-\epsilon)$), ($\alpha(r_0)$, $\beta(r_0)$), ($\alpha(r_0+\epsilon)$, $\beta(r_0+\epsilon)$). After a simple algebraic calculation, we obtain the radius $R(r_0, \epsilon)$ of the circle passing these three points. Further, taking the limit $\epsilon\rightarrow0$, we finally get the curvature radius of the shadow parameterized by $r_0$. For the KN black hole, we have
\begin{eqnarray}
 R=\frac{64\left(M r_0^3-Q^2
   r_0^2-a^2 M r_0 \cos
   ^2\theta _0\right)^{3/2} \left(r_0^3-3 M
   r_0^2+3 M^2 r_0-M \left(a^2+Q^2\right)\right) }{\left(r_0-M\right)^3
   \left(8 r_0^3 \left(3 M
   r_0-4 Q^2\right)-a^4 M \cos \left(4 \theta_0\right)-3 a^4
   M-4 a^2 M \left(a^2+6 r_0^2\right) \cos \left(2
   \theta_0\right)-24 a^2 M r_0^2\right)}.
\end{eqnarray}
When the black hole charge $Q$=0, $R$ will recover the result (\ref{Rad00}) for the Kerr black hole. We show the behaviors of the curvature radius $R$ in Fig. \ref{pKNRVQ} for the KN black hole. In Fig. \ref{KNRVa}, we set $\theta_0$=$\frac{\pi}{2}$ and $Q/M$=0.5. It is clear that the range of $r_0$ widens with the increase of black hole spin $a/M$ from top to bottom. $R_{\rm max}$ increases with $R_{\rm min}$ decreases with it. However when we take $\theta_0$=$\frac{\pi}{2}$ and $a/M$=0.5 shown in Fig. \ref{KNRVQ}, $R$ has a tiny difference. With the increase of $Q/M$, the range of $r_0$ is shifted to the left. $R_{\rm min}$ is found to decreases with $Q/M$ monotonously. However $R_{\rm max}$ shows a non-monotonic behavior. This is a novel phenomenon for the charged black hole.

%%%%%%%%%%%%%%%%%%%%%%%%%%%
%%%%%%%%%%%%%%%%%%%%%%%%%%%

%%%%%%%%%%%%%%%%%%%%%%%%%%%
\begin{figure}
\center{\subfigure[]{\label{KNConP2}
\includegraphics[width=7cm]{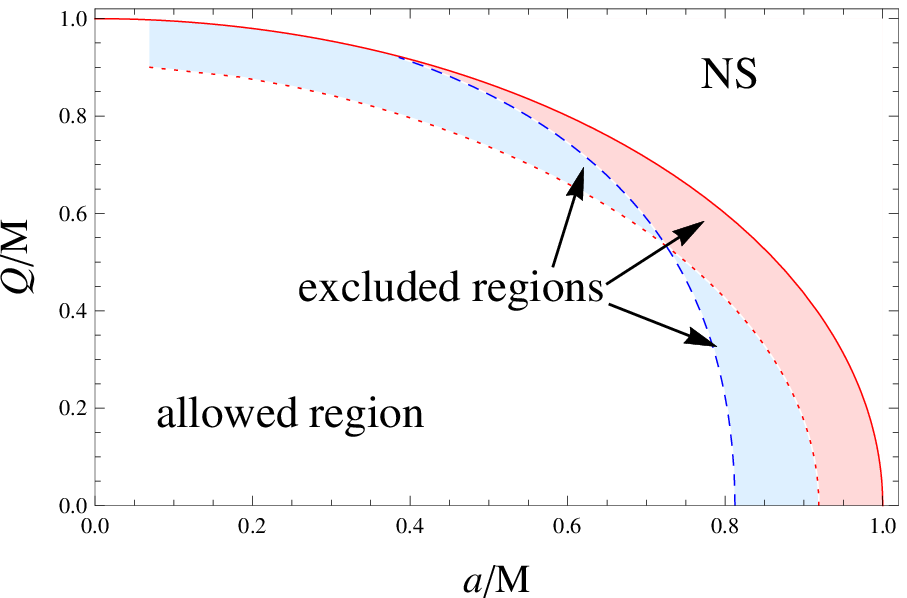}}
\subfigure[]{\label{KNConP3}
\includegraphics[width=7cm]{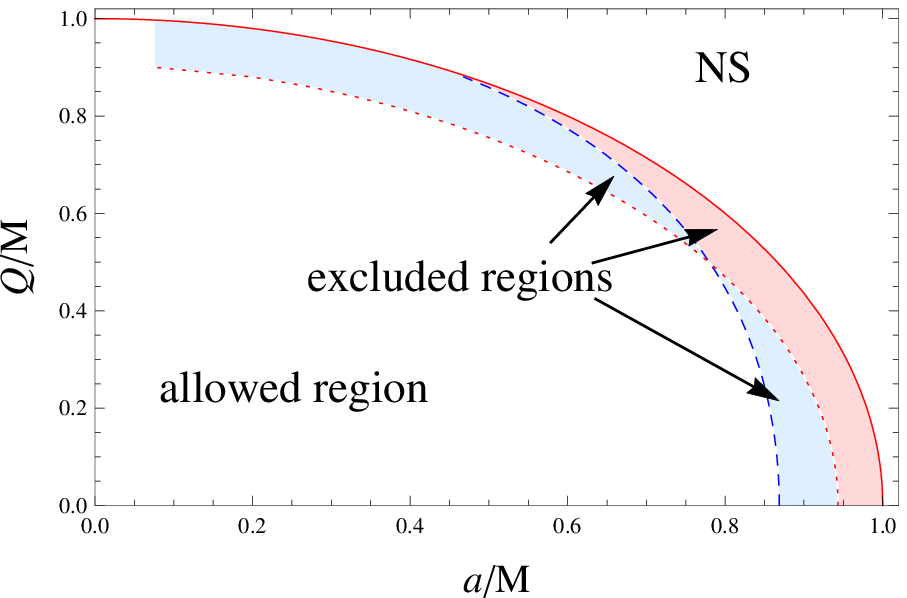}}}
\caption{$R_{\rm max}$ (blue dashed curves) and $R_{\rm min}$ (red dot curves) in the $Q/M$-$a/M$ plane. Red solid curves are for the extremal black hole curves, above which are the naked singularity regions denoting with ``NS". Shadow regions are for the excluded regions, and others are the allowed regions. (a) $\theta_{0}$=$\frac{\pi}{2}$. (b) $\theta_{0}$=$\frac{\pi}{3}$.}\label{pKNConP3}
\end{figure}
%%%%%%%%%%%%%%%%%%%%%%%%%%%

Adopting these constraints (\ref{cons2}), we calculate $R_{\rm min}$ and $R_{\rm max}$ and show the results in Fig. \ref{pKNConP3}. Red curves are for the extremal black hole curves, above which are for the naked singularity (NS), and bellow which are the black hole regions. $R_{\rm min}$ and $R_{\rm max}$ are described by the red dot and blue dashed curves, respectively. For each $\theta_0$, we find there are three shadow regions, respectively, corresponding to $R_{\rm max}>6.08$, $R_{\rm min}<4.31$, and both of them. These denote the excluded regions. Since they are near the extremal black hole curves, these excluded black holes are the near extremal ones. While these far away from the extremal black holes are allowed and can pass the constraints. With the decrease of $\theta_0$, we observe that $R_{\rm min}$ and $R_{\rm max}$ are slightly shifted toward to high black hole spin for small charge.

%%%%%%%%%%%%%%%%%%%%%%%%%%%
\begin{figure}
\center{\subfigure[]{\label{KNCa50}
\includegraphics[width=7cm]{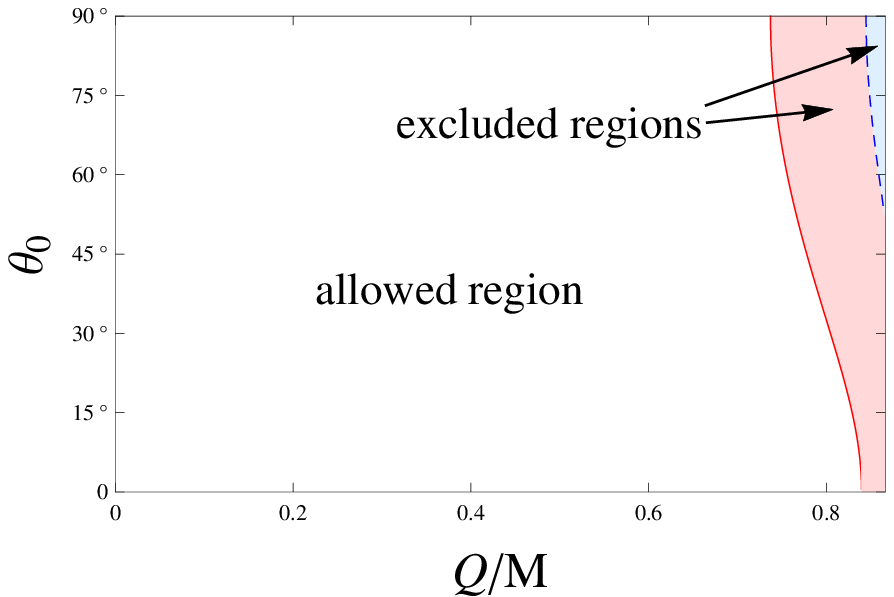}}
\subfigure[]{\label{KNCa95}
\includegraphics[width=7cm]{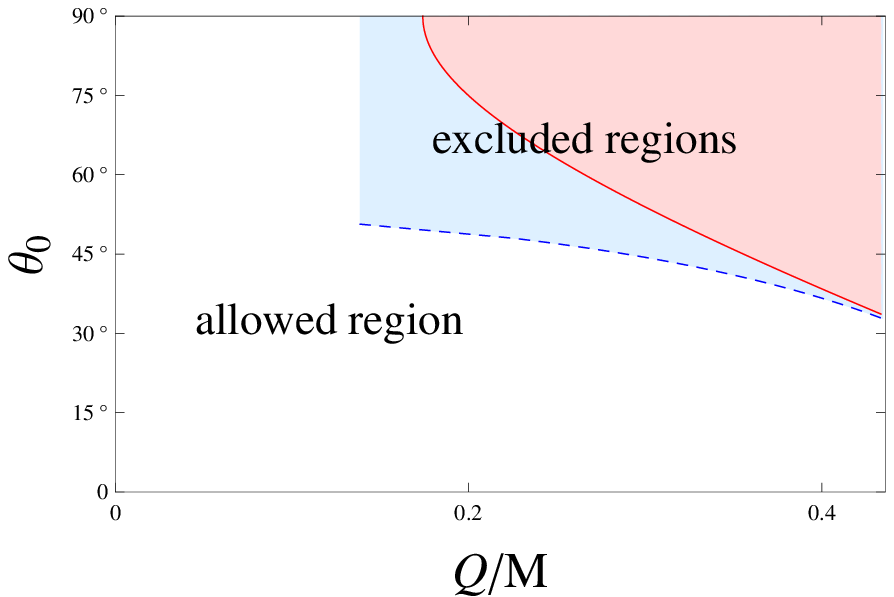}}}
\caption{Excluded and allowed regions in $\theta_0$-$Q/M$ plane. Blue dashed curves are for $R_{\rm max}$ and red solid curves are for $R_{\rm min}$. (a) $a/M$=0.5 with $Q_{\rm max}/M=0.8660$. (b) $a/M$=0.9 with $Q_{\rm max}/M=0.4359$.}\label{pKNCa95}
\end{figure}
%%%%%%%%%%%%%%%%%%%%%%%%%%%

In order to consider the influence of $\theta_0$ on the constraints, we show two characteristic constrained patterns for low and high black hole spins $a/M$=0.5 and 0.9 in Fig. \ref{pKNCa95}. For low spin $a/M$=0.5, the characteristic constrained pattern is described in Fig. \ref{KNCa50}. $R_{\rm min}$ will produce a stronger constraint than $R_{\rm max}$. Near extremal black holes will be excluded. The constraint becomes weak for low $\theta_0$. Nevertheless, the extremal black holes are always excluded. For high spin $a/M$=0.9, see Fig. \ref{KNCa95}, $R_{\rm max}$ dominate the constraint. The case with large values of $Q$ and $\theta_0$ will be excluded. Different from the $a/M$=0.5, the extremal black hole can survive for small $\theta_0$. On the other hand, small charged black holes can always pass the constraints as expected.

In summary, we calculate the curvature radius and place a constraint on the charge and spin for the KN black holes. High charged black holes are excluded via the constraints on $R_{\rm min}$ and $R_{\rm max}$. However, for low $\theta_0$, most black holes can pass the constraints.

\section{Summary}
\label{Conclusion}

In this paper, we constrained the spin and charge for the Kerr and KN black holes via the local curvature radius by the observation of the shadow from the M87*.

Adopting the results of EHT collaboration \cite{Kocherlakota}, we suggested constraints on the minimum and maximum curvatures are
\begin{eqnarray}
 4.31M\leq R_{\rm min}, R_{\rm max} \leq6.08M.\nonumber
\end{eqnarray}
Employing it, we constrained the parameter spaces for the Kerr and KN black holes.

For the Kerr black hole, we found that if the observer is near the equatorial plane, high spin black holes will be excluded. However for $\theta_0<31^{\circ}$, no constraint exists for the black hole spin. So higher $a/M$ and $\theta_0$ will be ruled out from M87*.

For the KN black hole, we firstly calculated its curvature radius. Then for different values of the parameters, $R_{\rm min}$ and $R_{\rm max}$ were obtained. For fixing $\theta_0$, we found that near extremal black holes are excluded, while the low charge and low spin black holes can pass the constraints. We also showed two characteristic constrained patterns for low and high black hole spins. For both the cases, high $a/M$ and $\theta_0$ are ruled out. For low spin, near extremal black holes are always excluded independent of $\theta_0$. However for high spin, the near extremal black hole can survive for small $\theta_0$. Despite that, low spin and small charge are always allowed.

In summary, we in this work adopt the curvature radius to constrain the black hole parameters from the observation of M87*. Since the curvature radius is a local concept, it provides us a novel point of view to test the black hole nature. Rotating black holes possess many other charges, such as the dilaton and axion, it is worth to generalize our approach to constrain these black hole hairs.

\section*{Acknowledgements}
This work was supported by the National Natural Science Foundation of China (Grants No. 12075103, No. 11675064, and No. U1931203), the 111 Project (Grant No. B20063), and the Fundamental Research Funds for the Central Universities (No. Lzujbky-2019-ct06).

\end{document}